\documentclass{article}
\usepackage{microtype}
\usepackage{graphicx}
\usepackage{subcaption}
\usepackage{booktabs} %

\usepackage{hyperref}

\usepackage[preprint]{icml2026}

\usepackage{amsmath}
\usepackage{amssymb}
\usepackage{mathtools}
\usepackage{amsthm}

\usepackage[dvipsnames]{xcolor}
\usepackage{algorithm}
\usepackage{algorithmic}
\usepackage[table]{xcolor}

\renewcommand{\algorithmiccomment}[1]{//#1}

\newcommand{\size}[1]{\lvert #1 \rvert}

\usepackage{rotating}

\newcommand{\BigO}[1]{\ensuremath{\operatorname{O}\!\left(#1\right)}}

\newcommand{\C}{\mathcal{C}}

\theoremstyle{plain}
\newtheorem{theorem}{Theorem}[section]

\newtheorem{lemma}[theorem]{Lemma}

\theoremstyle{definition}

\newtheorem{assumption}[theorem]{Assumption}
\theoremstyle{remark}
\newtheorem{remark}[theorem]{Remark}

\newcommand{\secref}[1]{Section~\ref{sec:#1}}

\newcommand{\appref}[1]{Appendix~\ref{app:#1}}
\newcommand{\appstworef}[2]{Appendices~\ref{app:#1} and~\ref{app:#2}}
\newcommand{\figref}[1]{Figure~\ref{fig:#1}}

\newcommand{\tabref}[1]{Table~\ref{tab:#1}}

\newcommand{\promptref}[1]{Prompt~\ref{pro:#1}}
\newcommand{\algref}[1]{Algorithm~\ref{alg:#1}}
\newcommand{\thmref}[1]{Theorem~\ref{thm:#1}}
\newcommand{\lemref}[1]{Lemma~\ref{lem:#1}}

\DeclareRobustCommand{\DE}[3]{#2}
\DeclareRobustCommand{\VAN}[3]{#2}

\icmltitlerunning{Algorithmically Establishing Trust in Evaluators}

\begin{document}

\title{Algorithmically Establishing Trust in Evaluators}

\twocolumn[
  \icmltitle{Algorithmically Establishing Trust in Evaluators}

  \begin{icmlauthorlist}
    \icmlauthor{Adrian de Wynter}{comp,yyy}
  \end{icmlauthorlist}

  \icmlaffiliation{yyy}{Microsoft}
  \icmlaffiliation{comp}{The University of York}

  \icmlcorrespondingauthor{Adrian de Wynter}{adewynter@microsoft.com}

  \icmlkeywords{evaluation, trust, LLMs-as-judges}
  \vskip 0.3in
]
\printAffiliationsAndNotice{}

\begin{abstract}
An evaluator, such as an LLM-as-a-judge, is trustworthy when there exists some agreed-upon way to measure its performance as a labeller. 
Traditional approaches either rely on testing the evaluator against references or assume that it `knows' somehow the correct labelling. 
Both approaches fail when references are unavailable: the former requires data, and the latter is an assumption, not evidence. 
To address this, we introduce the `No-Data Algorithm', which provably establishes trust in an evaluator without requiring any labelled data. 
Our algorithm works by successively posing challenges to said evaluator. 
We prove that after $r$ challenge rounds, it accepts an evaluator which knows the correct labels with probability $ \geq 1 - (1/4)^r$, and reliably flags untrustworthy ones. 
We present formal proofs of correctness, empirical tests, and applications to assessing trust in LLMs-as-judges for low-resource language labelling. 
Our work enables scientifically-grounded evaluator trust in low-data domains, addressing a critical bottleneck for scalable, trustworthy LLM deployment.

\end{abstract}

\section{Introduction}\label{sec:intro}
One of the most fundamental problems in AI is the question of measurement. 
Many commonly used metrics have a certain level of unreliability; for example, by not corresponding to human judgements \citep{reiter-2018-structured,gehrmann2023repairing,bavaresco2024llmsinsteadhumanjudges,LLMLXEval}. 
With the rise of LLMs as evaluators, typically known as LLMs-as-judges, scaling measurement to multiple subproblems has become ubiquitous \citep{de-wynter-2025-awes}. 
However, their reliability and trustworthiness as measurement tools has led to division within the research community, 
specifically, about the validity of some results \citep{gehrmann2023repairing,rabinovich-anaby-tavor-2025-robustness,moghe-etal-2023-extrinsic}. 

A key issue on the question of measurement is whether the evaluation tool (the \textit{evaluator}) may be trusted. 
This trust is typically established with a ground truth artifact, such as a labelled dataset for statistical analysis, or a benchmark. %
However, references may not exist or be scarce, due to annotation cost or uniqueness of the problem. 
In this case, statistical conclusions would not be meaningful. 
Even when references exist, contamination of benchmarks has become a major concern (see, e.g., \citealt{sainz-etal-2023-nlp}). 
Hence, when no references exist, scientifically establishing trust on an evaluator can only be done with a formal proof.

We introduce an algorithm (the `No-Data Algorithm') which allows trust assessment \textit{without} any reference data. 
It is designed for contemporary settings, such as LLMs-as-judges, where an \textit{evaluator} claims knowledge of the correct labels, and constrained to a binary classification setting. 
A \textit{verifier} then decides, based on the evaluator's successful responses to specific challenges, whether it is trustworthy. 
This requires a linear number of queries, no labelled datasets (e.g., development or training sets), or prior knowledge of the labels. 
The method is inspired by zero-knowledge proof protocols, such as two-factor authentication. 

\subsection{Contributions}

Our contributions are theoretical and experimental. 
Theoretically, we prove that after $r$ calls, the No-Data Algorithm establishes trustworthiness on the evaluator to a $(1/4)^{r}$ probability of error. 
This measure robustly reflects whether the evaluator actually has label knowledge. 
A knowledgeable evaluator will have high accuracy and success rate. 
Conversely, a lying evaluator\footnote{In this work we use `lying' to denote both uninformed and sycophantic evaluators, and we evaluate both cases.} may achieve high accuracy only by chance, but will not attain a high success rate. 

Empirically, we test the No-Data Algorithm with traditional machine-learning (i.e., a decision tree) and contemporary (r. LLMs) evaluators. 
We demonstrate an application to assessing an LLM-as-a-judge's trustworthiness in a novel dataset in West Frisian, a low-resource language. 
We also include ablation studies with LLMs-as-judges, prompts, algorithm components, and extend our approach to $k$-ary label sets.\footnote{All data is in \url{https://github.com/adewynter/no_data_algorithm}.} 

Our work shows that trust in an evaluator may be established formally without references, enabling scientific evaluation in domains with extremely scarce labels, such as market research, low-resource languages (the subject of our experiments), medicine. 
To our knowledge, ours is the first mathematically rigorous approach to establishing trust in an LLM-as-a-judge without labelled references.

\section{Related Work}\label{sec:relatedwork}

There has been considerable work on attempting to create reliable evaluators. Most of contemporary work relies on LLMs and various prompting strategies or call stacks, although the problem of (un)reliability/trustworthiness of an evaluator is older than generative models. 
For example, it has been known that the use of BLEU for tasks other than English-based machine translation is not accurate \citep{liu-etal-2016-evaluate,novikova-etal-2017-need}; and even then, that it does not correspond well to human judgements \citep{reiter-2018-structured,gehrmann2023repairing}. 

More generally, it is known that metric choice, and hence the choice of evaluator, is system-dependent \citep{chen-etal-2024-humans,moghe-etal-2023-extrinsic,flamich2025feedbirdsscoreaccuracynaturalness,vondäniken2024measuredependenceautomatedmetrics,pombal2025addingchocolatemintmitigating}. 
Poorly-developed rubrics also lead to unreliable human judgements \citep{clark-etal-2021-thats,van-der-lee-etal-2019-best}, and thus the evaluator trustworthiness problem is also a rubric problem. 
There are algorithms \citep{northcutt2017rankpruning,NEURIPS2018_f2925f97} and paradigms \citep{confidentlearning} for learning with noisy teachers. 
Likewise, provable methods have been used with success in other areas, such as representation learning \citep{jovanovic2023fare}, and to improve robustness via synthetic data generation \citep{dimitrov2022provably,Fischer2022}. 
However, they all assume that a (perhaps noisily) labelled dataset exists.

Relevant to our work, the use of LLMs as evaluators has promised extraordinary scaling capabilities, such as fast data generation and evaluation without humans \citep{GPT4,chiang2024chatbotarenaopenplatform,chiang-lee-2023-large,liu-etal-2023-g,rethinkingsemantic,NEURIPS2023_91f18a12}. 
Some of these criticisms are that they do not correlate with human judgements well \citep{bavaresco2024llmsinsteadhumanjudges,LLMLXEval,chen-etal-2024-humans,rtplx,10.1162/tacla00685}; 
are sensitive to their prompt \citep{lu-etal-2022-fantastically,hida2024socialbiasevaluationlarge,ye2022the}; 
their performance could be the product of memorisation \citep{PlagiariseLee,dewynter2023evaluation,sainz-etal-2023-nlp}; 
and even that they do not understand the task at all \citep{dewynter2023i,webson-pavlick-2022-prompt,ye2022the}. 
Their reasoning capabilities have been put into question, from the results being strongly dependent on the choice of metrics \citep{schaeffer2023are,chen-etal-2024-humans}, to findings that their output reasoning steps contain spurious reasoning \citep{TurpinCoT,lanham2023measuringfaithfulnesschainofthoughtreasoning,wu-etal-2024-decot}. 
Most of the work on improving trust, however, relies on prompting strategies or call stacks to other LLMs \citep{li2024llmsasjudgescomprehensivesurveyllmbased}. 

The main subroutine from the No-Data Algorithm, which we call the Evaluator-Verifier (EV) protocol, is based off the well-known Arthur-Merlin (AM) protocol from \citet{babai}. It is a type of zero-knowledge proof; see \citet{goldreich} and \citet{arorabarak} for primers on the subject. 
That said, the EV protocol is different in many regards to the AM protocol, since it is designed to fit arbitrary inputs and the larger No-Data Algorithm. 
In particular, it does not provide privacy or secrecy as other cryptographic protocols do; instead, we adapt the challenge-response game to evaluation. 
Likewise, our use of the rubric and the aggregator functions are generalisations to partial functions that are not present (or needed) in previous works.

\section{Definitions and Problem Setup}\label{sec:definitions}

\subsection{Notation}

Our work relates to \textit{datapoints} $x\in X$ and \textit{labels} $y \in Y$. 
Whenever the label set $Y$ is unspecified, it is the binary label set $Y = \{0, 1\}$. 
We also let $X \subset \{0, 1\}^n$; that is, every datapoint is a $n$-bit binary string modelling a phenomenon. 
The assumptions on $X$ and $Y$ are to simplify our proofs; see \secref{discussion} for a discussion on their relevancy. Experimentally, $X$ may be other (concrete) data types, such as natural language strings. 
For an $x \in \{0, 1\}^n$, we denote its \textit{relevant subset set} as $S_x \subset 2^x$. The construction of this specific set will depend on the setup, and will be discussed in detail in \secref{pvpdefinition}.

For two functions $f \colon B \rightarrow C$, $g \colon A \rightarrow B$ we write the composition $f\circ g \colon A \rightarrow C$ as $fg$. 
For two sets $A$, $B$, we denote equality \textbf{up to isomorphism} as $A \cong B$, where $A \cong B$ if $f(A) = f(B)$ but $A \neq B$ for some $f \colon U \rightarrow V$. 
We also consider the special case of equality \textbf{up to permutation} over subsets as $A \equiv B$, where $A \equiv B$ if $\forall a \in S_A, \exists b \in S_{B}\,.\,g(a) = g(b)$, for some fixed function $g \colon U \rightarrow V$. 
We assume all functions to be deterministic throughout, unless stated otherwise.

\subsection{Definitions}

In line with standard learning theory \citep{Valiant}, we assume that there exists an \textit{unknown} map $f \colon X \rightarrow Y$ taking datapoints to labels faithfully--that is, $f$ provides the true labelling. %
Our task is to determine whether a function $E \colon X \rightarrow Y$ (a labeller, or evaluator) is equivalent to $f$, \textit{without} any knowledge of the map itself. This is measured with its error rate, or, conversely, its accuracy. 

We also introduce the following definitions:

A \textbf{criterion} $c \colon X \rightarrow \{0, 1\}$ classifies datapoints. 

A \textbf{rubric} $C \colon \times_{i, \dots, n} \{0, 1\} \rightarrow \{0, 1\}^n$ maps the binary outputs from an ordered set of criteria $\C = \{c_1, \dots, c_n\}$ to a bitstring which is the concatenation of these outputs: $C = c_1|c_2|\dots |c_n$. 
We say \textbf{a rubric is total} if it explicitly decomposes nonlinear criteria with arity $> 1$ (e.g. the xor $c_i = c_a \oplus c_b$), and write it as $\bar{\C}$. 
When the evaluation of said criteria may be done separately (e.g., by testing first $c_a$, then $c_b$, and then their xor $c_a \oplus c_b$), we say it is a \textbf{total evaluation}. 

The \textbf{aggregator} $\sigma \colon \{0, 1\}^n \rightarrow Y$ is a function that maps $n$-bit binary strings (rubric outputs) to $Y$. 

For two $x, x' \in X$, we say that \textbf{$x$ is similar to $x'$} if both $x \cong x'$ and $x \equiv x'$. 

Finally, in line with the AM protocol, we use \textbf{lying} to refer to any evaluator that presents unsubstantiated claims of correctness, including both uninformed (random guessing) and sycophantic (overconfident without knowledge) behaviour. 
The difference is that the latter could perform well in-distribution, but fail otherwise. The former, on the other hand, will simply not know how to solve the problem and fail at both. 
Without labels it is difficult to discern both cases, although we show that our algorithm is capable of it. 

Remark that, intuitively, the rubric is the set of reasons \textit{why} a datapoint belongs to a certain label class (e.g, an annotation rubric passed in to human labellers). 
The label is thus based on evaluating the rubric and aggregating its criteria with $\sigma$ (e.g., with a majority vote). 
Natural-language examples are in \appref{rubrics}; 
experiment with potential rubric weaknesses in \secref{lowres}; 
and discuss them further in \secref{discussion}. 

We are now ready to introduce our main assumption: 
\begin{assumption}
The map $f \colon X \rightarrow Y$ is a composition of the rubric and the aggregator: 
\begin{equation}
    f = \sigma C.
\end{equation}
\end{assumption}
This is not a strong assumption: after all, datapoints must have a reason (criterion) to acquire a label. 
This reason must be clearly stated; and, when there is more than one reason, there must be a way to decide how to aggregate them. 
In realistic scenarios, $C$ or $\sigma$ may not be known. There, our method may still be applied with user-defined surrogates, but derived trust bounds are with respect to these surrogates rather than any `true' but unknown ground truth. 
We come back to this in \secref{discussion}.

\section{The No-Data Algorithm}\label{sec:nodatalagorithm}

The No-Data Algorithm is an algorithm with two components, or players: the \textbf{evaluator} and the \textbf{verifier}. %
Throughout the algorithm's run, for every datapoint $x \in X$, the evaluator must convince the verifier that its choice of label for $x$ is correct (i.e., that it knows $f$). 

For this, the players run a multi-round sub-game which we refer to as the \textbf{Evaluator-Verifier} (EV) \textbf{protocol}. 
The EV protocol takes in $x$, and either succeeds or fails, w.h.p., based on challenges posed by the verifier on a $x'$ similar to $x$ generated by the evaluator. 
The EV protocol returns the proposed label and the status (success or failure). 

The No-Data Algorithm then flips the label with probability $\phi$ in case of failure. 
After observing all $X$, it returns the success count and final labelling. 
The evaluator's trustworthiness (and hence the final labelling's) is established via the average success rate.

The full algorithm is in \algref{algorithm1}. 
We provide proofs of correctness in \secref{correctness}.

\begin{algorithm}[h]
\begin{algorithmic}[1]
   \STATE {\bfseries Input:} Unlabelled data $X$, evaluator $E$, verifier $V$, criterion $C$, flip probability $\phi$, rounds $r$
   \STATE $predictions \gets \{\}$
   \STATE $successes \gets \{\}$
   \FOR{$x \in X$} \label{lst:line:startfor}
   \STATE $success, y \gets \textsc{EV}(x, E, V, C, r)$\hfill\COMMENT{ Run for $r$ rounds}
     \IF{$success$} %
       \STATE $predictions \gets predictions \cup \{y\}$ \label{lst:line:correctstatement}
       \STATE $successes \gets successes \cup \{1\}$
    \ELSE
        \STATE $\tilde{y} \gets \lnot y \text{ with probability } \phi$ \hfill\COMMENT{ The opposite label} 
        \STATE $predictions \gets predictions \cup \{\tilde{y}\}$
        \STATE $successes \gets successes \cup \{0\}$
     \ENDIF
   \ENDFOR \label{lst:line:endfor}
   \STATE Return $predictions$, $successes$
\end{algorithmic}
\caption{The No-Data Algorithm. For every $x \in X$ it calls the EV protocol to determine the trustworthiness of the label returned by the evaluator, $y = E(x)$. 
Based on its output, the algorithm flips $y$ w.p. $\phi$ in case of failure. Finally, it returns the success count and the labels.}\label{alg:algorithm1}
\end{algorithm}

\subsection{The Evaluator-Verifier (EV) Protocol}\label{sec:pvpdefinition}

In the EV protocol both players are given a datapoint $x\in X$ and rubric $C$. The aggregator $\sigma$ is not given. 
The evaluator \textit{claims} to know $f \colon X \rightarrow Y$; and, since $C$ is known, the claim reduces to the evaluator having knowledge of $\sigma$. 
The verifier's goal is thus to be sufficiently sure that the evaluator's labelling of a given $x \in X$ may be trusted. 

\subsubsection{Setup}\label{sec:setupev}

\begin{figure}[h]
    \centering
    \includegraphics[width = 0.85\linewidth]{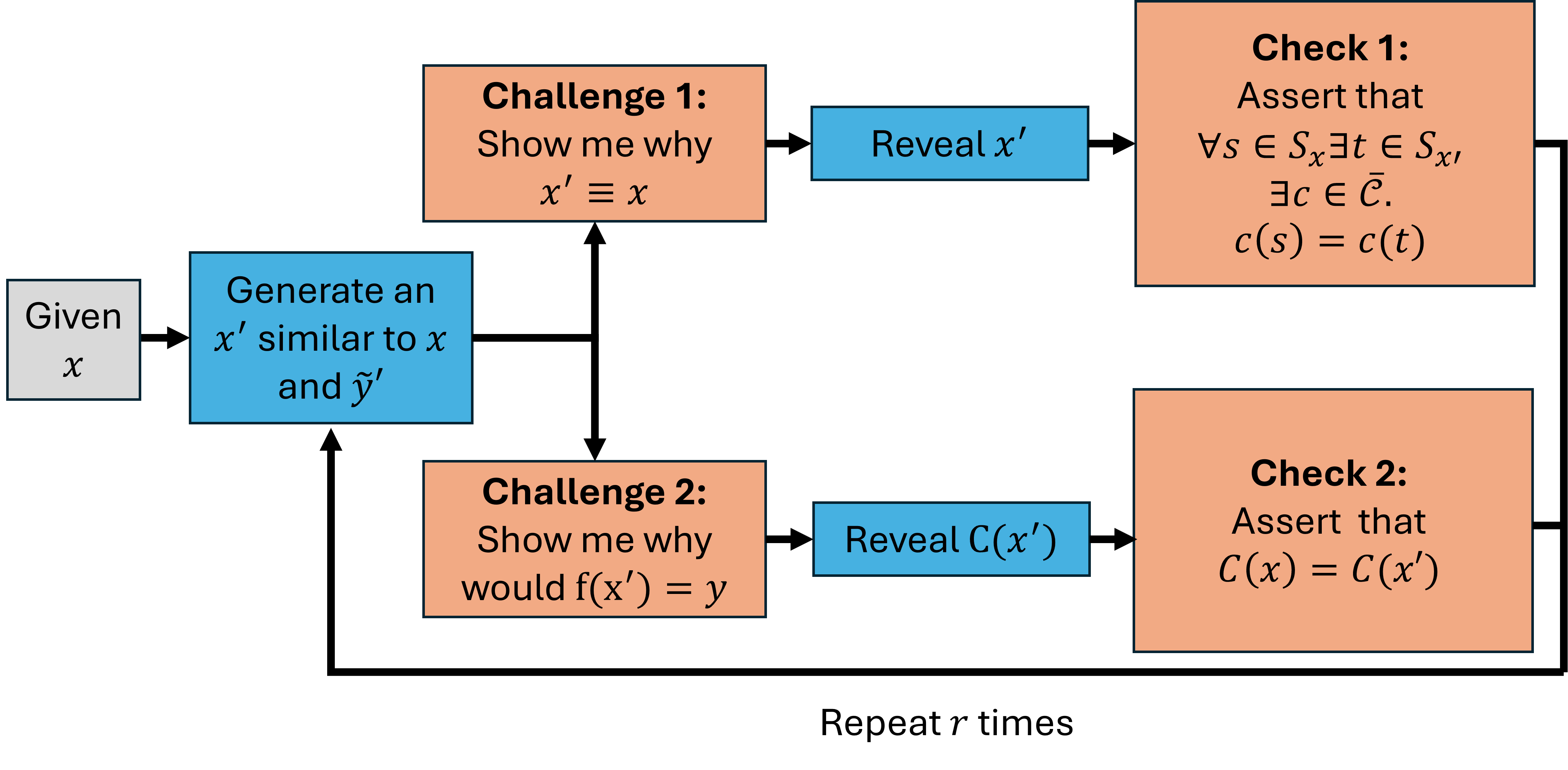}
    \caption{EV protocol flow. 
    At every round, the evaluator (blue) generates an $x'$ similar to $x$, and a partial label $\tilde{y}'$. 
    It then answers one of two (chosen uniformly at random) challenges by the verifier (orange). 
    If the evaluator does not pass the challenge, the protocol returns failure. Otherwise, the game is repeated. 
    If the rounds are over, it returns a succeeded state. In either case, it also returns $\tilde{y}'$.}
    \label{fig:evpimg}
\end{figure}

The EV protocol is played for $r$ rounds. 
At every round, the evaluator generates a similar datapoint $x'$ to $x$, and a `partial' label $\tilde{y}'$ that--supposedly--is equal to $y$.\footnote{Remark that the process by which the evaluator comes up with $x'$ does not need to be generation in the LLM sense. It is instead left up to the evaluator, not unlike other zero-knowledge proofs.} %
It is partial because there is no access to $f$, and hence the best the evaluator can do is to guess $\tilde{y}'$ based on $C(x')$. 
Once it has generated the tuple $\langle x', \tilde{y}'\rangle$, the evaluator may not change it in this round. 
The verifier poses one of two challenges, selected uniformly at random. 
Both challenges measure equality between $x'$ and $x$, either up to isomorphism (validating the label) or up to permutation (structurally validating the criteria). 
If the evaluator fails, the protocol terminates and returns failure. 
Otherwise, it starts over. 
Regardless of the outcome, the protocol also returns $\tilde{y}'$. 
See \figref{evpimg} for a diagram of the protocol.

\paragraph{Challenge 1 (Equality up to Permutation)}\hfill\\
\textbf{Setup:} The verifier tests the evaluator by asking it to show a similar point that leads to the same total evaluation of $C$, namely, whether $x \equiv x'$ over $\bar{\C}$. 

\textbf{Check:} The verifier asserts that
\begin{align}
    \forall s \in S_x, \exists t \in S_{x'} \,.\, \forall c \in \bar{\C},\,c(s) = c(t);%
\end{align}

that is, if there is at least one substring in $x'$ matching the (total) evaluation of a criterion from $x$. 

\paragraph{Challenge 2 (Equality up to Isomorphism)}\hfill\\
\textbf{Setup:} The verifier asks why $f(x') = y$. Since $y$ is not available, it checks whether $x' \cong x$. 

\textbf{Check:} The verifier asserts that $C(x') = C(x)$.

\begin{remark}[]
Since the challenges are public, the evaluator could cheat and generate an $x'$ to pass one of the challenges, \underline{but not both}. 
Otherwise, the implication is that it knows $f$, and it will pass the challenges every time. 
This is why the challenges assert \underline{both} structure ($x \equiv x'$) and final encoding ($x \cong x'$). 
This observation is central to our correctness proofs from \secref{correctness}. 
\end{remark}

\section{Correctness}\label{sec:correctness}

\begin{lemma}[EV Protocol Correctness Bound]\label{lem:pvpcorrectness}
The probability that the verifier fails to detect a lie by the evaluator in the the EV protocol sub-game after $r$ rounds is $(1/4)^r$. 
\end{lemma}
\begin{proof}
(Sketch; a complete proof is in \appref{pvpcorrectnessfull}). 
We split the proof in two: (1) whether the challenges are necessary and sufficient to detect if the evaluator is lying; and (2) the probability bound to which this detection occurs. 

For the first part, note that Challenge 1 determines the internal structure of the string, but does not account for the valuation of $C(x')$. 
This challenge is necessary because two $x, x'$ could lead to the same evaluation of some $c \in \C$ (e.g., disjunctive clauses), yet this does not imply that they are equal up to permutation. 

However, Challenge 2 requires strict equality in $C(x') = C(x)$. It is indeed order-dependent, but does not account for internal structure. 
Hence both challenges provide some, but not all, information about $x, x'$ and the claimed $f(x), f(x')$. 
Thus they are together sufficient and necessary to determine if an evaluator is lying. 

Since they are mutually exclusive in the information they provide, it follows that an evaluator passing both challenges will \textit{not} be lying. 
Thus, a lying evaluator may only be able to pass one challenge. 
Hence we handle two cases where the evaluator could lie (or cheat): either (1) it can generate an appropriate $x'$, but not a correct $C(x')$; or (2) it provides a correct $C(x')$ but an incorrect $x'$. 
In either case, the probability that it gets lucky and/or passes the challenge by cheating is $1/4$; after $r$ rounds, it is $(1/4)^r$. 
\end{proof}

\begin{theorem}[No-Data Algorithm Correctness Bound]\label{thm:ndacorrectness}
Suppose an evaluator has accuracy $\alpha$ on a balanced dataset $X$ of size $n$ with binary labels $Y = \{0, 1\}$. 
Assume the evaluator always lies on any $x \in X$ iff it incorrectly labels the datapoint. 
Then if the EV protocol is ran for $r$ rounds, the expected accuracy of the No-Data Algorithm is given by
\begin{equation}\label{eq:ndaerrorbound}
\mathrm{E}[\text{correct}] \leq 1 - (1 - \alpha) \left(1 - \phi+ \phi\left(\frac{1}{4}\right)^r\right),
\end{equation}
for a chosen value of $\phi$ and over the same dataset $X$. 
\end{theorem}
\begin{proof}
(Sketch; a complete proof is in \appref{ndacorrectnessfull}).  
The accuracy of any evaluator is directly tied to its success in the EV protocol: if it knows the label, then the checks will pass and $\tilde{y}' = y$. 
From \lemref{pvpcorrectness}, this holds with probability $1 - (1/4)^r$. 
An application of the union bound yields the desired result.%
\end{proof}

\thmref{ndacorrectness} implies that a good evaluator with relatively high accuracy and success rate, and a well-calibrated $\phi$, will have a true accuracy higher than what the No-Data Algorithm outputs. 
Hence, reliable evaluators, as per the algorithm, may be used with the expectation that their `true' performance will be better. 
Another interpretation of this result is that there is no way to recover unknowable labels when the evaluator is unable to convince the verifier. %

Remark that the assumption for \thmref{ndacorrectness} for the evaluator always lying when mislabelling (and hence always being wrong) is not necessarily applicable to all evaluators. A particularly tricky evaluator could still guess the right label, yet not pass the challenges. 
We consider this out of scope, since the ability to guess a label without explaining how is not a viable way to establish trust, especially in scenarios without any references.

\section{Experiments}\label{sec:experiments}

We present two experiments: an empirical test of the theory (\secref{dt}), and an application to a realistic scenario (\secref{lowres}). 
Ablation experiments and extensions are in \appstworef{kary}{fullresults}. 
We used two corpora: one which the evaluators know how to label (\textit{in-phenomenon}, or IP), and one they do not (\textit{out-of-phenomenon}; OOP). 
This naming is to highlight that the measurement is altered by changing the dataset--semantically we are modelling different phenomena (i.e., distinct $f \colon X \rightarrow Y$). 
We refer to the hypothetical setup where we know the labels as \textit{knowable} (r. \textit{unknowable}). 
Knowable setups are only to illustrate how the algorithm's labelling adheres to the `true' predictions. In practice, they would be unavailable. 
We run the No-Data Algorithm with $r = 3$. 
Remark that algorithmic baselines for this scenario are challenging, since approaches such as ensembling do not provide formal bounds of correctness, nor account for lying evaluators. 
See \appref{baseline} for a baseline analysis, and \appref{experiments} for prompts and methodology details. 

\subsection{Experiment 1: Synthetic Data without Natural Language}\label{sec:dt}

In this experiment we aim to understand the behaviour of the No-Data Algorithm in a setup close to that of the theory (binary strings). 
For this, we created two disjoint synthetic datasets over $\{0, 1\}$. 
This allows us to ablate any potential memorisation concerns, and to control the rubrics. The rubrics are in \tabref{criteriatable}. %
For both OOP and IP, $\sigma$ is the majority vote. 
The test sets for both IP and OOP included 498 entries and are balanced. 

\subsubsection{Setup: Evaluators and Verifiers}\label{sec:evalsetup}

\textbf{Evaluators:} We used a decision tree (DT), and the highest-performing LLM we tested (o3-mini; \citealt{OpenAIo3mini}).\footnote{The full results, in \appref{fullresults}, further include another closed- and two open-source models.} 
Both observed a balanced training split of the IP dataset. %
That is, we trained the DT to ensure that the evaluator learnt both IP's $\sigma$ and $C$, even though during generation it only had access to $C$, and not $\sigma$. 
The LLM, as `training', observed the rubric in the prompt, plus five exemplars, and was made aware of the aggregation function \textit{only} for the labelling step. 
For generation it was asked to pick the best candidate from the (training) dataset instead of outputting a new point, as we observed marked performance decreases otherwise (\appref{fullresults}). %
During OOP, the LLM was given the rubric for IP and an OOP datapoint. %
We tuned $\phi$ to $\phi = 0.4, 0.1$ for the DT and the LLM, respectively. 
These are the worst-case scenarios, since, as per \thmref{ndacorrectness}, it is preferable to have $\phi$ close to the knowable accuracy.
\\

\textbf{Verifier:} We used a rule-matching algorithm matching the test set to its appropriate rubric. %

\captionsetup[table]{name=Rubric}
\setcounter{table}{0}
\begin{table*}[ht]
\centering
  \caption{Rubrics and criteria for our first experiment. Note that $c_1$ for the IP rubric is an exclusive-or operation: while this rubric contains three criteria, the total rubric contains five: $\{c_0, c_2, c_{1, a}, c_{1, b}, c_{1}\}$; 
  where $c_{1,a}, c_{1,b}$ are the clauses for $c_{1}$, and $c_{1}$ is the exclusive-or itself. 
  }
  \begin{tabular}{ll}
    \textbf{Rubric} & \textbf{Definition} \\
    \toprule %
    \textbf{In-Phenomenon} & \\
    $c_0$ & $y=1$ if $x$ has an even number of ones; else $0$ \\
    $c_1$ & $y=1$ if $x$ starts with a zero or contains $10101$, but not both; else $0$ \\
    $c_2$ & $y=1$ if $x$ has strictly more than five ones; else $0$\\ 
    \midrule
    \textbf{Out-of-Phenomenon} & \\
    $c_0$ & $y=1$ if $x$ contains the substring $111$; else $0$\\
    $c_1$ & $y=1$ if $x$ ends with a one; else $0$\\
    $c_2$ & $y=1$ if $x$ contains the substring $110001$; else $0$\\
    \bottomrule
  \end{tabular}
  \label{tab:criteriatable}
\end{table*}

\subsubsection{Results}\label{sec:results}
The results are in \tabref{resultstable}. 
The DT and the LLM adjusted to the predictions: 
accuracies with the No-Data Algorithm were within $-2$ and $0$\% of the knowable case in both IP and OOP. 
The success rate differences between IP (100\% for DT; 81\% for the LLM) and OOP (r. 5\% and 28\%) provided better information as to which setup was the one the evaluator truly knew. 

We also performed ablation studies on the relationship of lying with evaluators and generators; the need for flipping the labels; and the effectiveness of generation strategies in various LLMs. 
Full details for all studies are available in \appref{fullresults}. 
We found that not flipping the labels led to lower accuracies in the No-Data Algorithm, especially when the evaluator lied and within IP. 
When the evaluator lied about knowing $\sigma$ or $f$, the success rates were lower in the IP case, but remained steady in OOP. 
These results align with the expectations from \secref{correctness}. 
The final study showed that LLM generation \textit{in this problem} was slightly lower-performing (-3\%) when compared to picking a string from the dataset. 
This is expected for non-natural language problems, does not impact the results, and does not extend to natural language. 
We discuss this further in \secref{discussion}. 

\begin{table*}[ht]
\centering
  \caption{Results for the \textit{known} (test score if the labels were known) and \textit{unknown} (output of the No-Data Algorithm) cases. 
  The known case is a reference that in practice does not exist. 
  The flip $\phi$ is set near the error rate in the known case ($0.4$ for the DT; $0.1$ for the LLM), since it is arguably the worst value it can have. 
  This is needed in this experiment because the generator acts as an oracle, convincing the verifier every time, and hence the No-Data algorithm reduces to simple evaluation (the known case). 
  However, the success rate is extremely low in DT OOP (unknown): even though there is no way for us to know whether the evaluator's score is trustworthy, the success rate indicates deception. 
  }
  \begin{tabular}{lcccc}
    & \textbf{DT} (known) & \textbf{DT} (unknown)  & \textbf{LLM} (known) & \textbf{LLM} (unknown) \\
    \toprule %
    \textbf{IP} & \\
    Successes/Flips  & --- / ---   & 100.0 / 0.0 & --- / --- & 81.3 / 1.8\\
    Accuracy/F$_1$   & 62.2 / 58.8 & 62.2 / 59.8 & 99.8 / 99.8 & 97.6 / 97.6 \\ 
    \midrule
    \textbf{OOP} & \\
     Successes/Flips & --- / ---   & 4.8 / 46.4 & --- / --- & 28.0 / 6.0 \\
     Accuracy/F$_1$  & 54.2 / 54.2 & 52.8 / 52.1& 60.6 / 66.2 & 59.0 / 64.0\\
    \bottomrule
  \end{tabular}
  \label{tab:resultstable}
\end{table*}

\subsection{Experiment 2: Low-Resource Language Labelling}\label{sec:lowres}

This experiment evaluates a setup much closer to a realistic scenario. 
The goal is to evaluate a chatbot's suitability for deployment in a low-resource language. 
Added to that, the rubric contains ambiguous criteria, and the criteria values are not known to the verifier or the evaluator. Both components are LLMs-as-judges. 

\subsubsection{Data}

We created a multi-domain dataset (1,015 entries total) with randomly-drawn prompts from OpenOrca \citep{OpenOrca}, MMLU \citep{hendryckstest2021}, OpenCode \citep{ahmad2025opencodereasoning}, and WildChat \citep{zhao2024wildchat}. 
The task is to judge whether the \textit{output} of an LLM (GPT-4o) to these prompts is correct based on a rubric. 
Hence, every entry in the corpus is a pair of prompts and outputs professionally translated into West Frisian and annotated by four native speakers. 
West Frisian is a language that, albeit spoken by half a million people, is within the class of `exceptionally limited resources' with `virtually no labelled data to use', as per the taxonomy from \citet{joshi-etal-2020-state}. 
See \secref{ethics} and \appref{experiments} for details on annotation. 

The rubric is written in natural language and consists of six criteria standard for chatbots deployed in user-facing applications, ranging from `the response must be in West Frisian' to `the response must be correct and without syntax errors'. 
For OOP we wrote a separate, almost-orthogonal rubric (e.g. `the response must be in Dutch', as the model often responded in Dutch), and included a compound entry (`one of B or C must be correct, otherwise the response is zero'), bringing the total OOP criteria to seven. 
Full rubrics are in \appref{rubrics}. 
The $\sigma$ for this corpus is 1 if all criteria are 1, otherwise 0.

The evaluator and verifier is GPT-4.1 \citep{gpt41}, although with different prompts, and thus different functions. 
We split the dataset into train (500) and test (515). 
Ahead of running the experiments, we calibrated the prompt strategies that worked best under the IP known case for both the final labels and the per-criterion breakdown. 
This is analogous to the setting where a user would tune their verifier prior to deployment on the unknown cases. 

This experiment is by design more challenging than the one from \secref{dt}. 
Here neither the verifier nor the evaluator know the label \textit{or} the criteria values: instead, both create their own label sets, and generate their own datapoints. 
The rubric has also ambiguity present (e.g., the IP rubric requests that the model must always answer `even if its answer is wrong'; but also that the response `must be correct'), and only the OOP rubric is decomposable. 
The ambiguity was unintentional, although not uncommon for realistic scenarios. 
We discuss this further in \secref{discussion}.

\subsubsection{Results}

The results are in \tabref{frisian}. The LLMs tested had high success rates and accuracies in the IP setting, and the converse in the OOP case. 
There were some nuanced differences when compared to the synthetic experiments from \secref{dt}. 
Namely, the difference in performance was higher when $\phi$ was low (up to -3.9\% and -2.2\% accuracy in IP and OOP, r.). Nonetheless, as expected, success rates remained steady regardless of value of $\phi$. 

To further probe the impact of LLM choice on the results we also tested an evaluator \textit{not} proficient in West Frisian (Qwen 2.5B VL 7B; \citealt{qwen2.5-VL}). 
This would be an evaluator which does not lie, but will be wrong regardless. 
We found that in the IP case, when the model was given the correct rubric, the success rate was still low, hovering at around 35\% in IP; and 2\% for OOP (where it did lie), for all values of $\phi$. 
The full results are in \appref{linguisticcompetency}. 
We also probed the effects of ternary label classification setup on the same dataset (\appref{kary}). 
Our findings largely adjusted to our other results, with the No-Data Algorithm spotting lying evaluators effectively.

\begin{table*}[ht]
    \centering
    \caption{Performance of the No-Data Algorithm with GPT-4.1 as evaluator and verifier, and the natural-language setup as dataset. 
    It can be seen that the performance of the evaluator adjusts to the predictions, with near-random accuracy in OOP and close-to-known accuracy in IP. 
    Remark that the success rate remains comparatively stable, regardless of the chosen value of $\phi$. Hence the most appropriate quantity to measure is the success rate, while $\phi$ may be tuned later for higher accuracy gains.
    }
    \begin{tabular}{l|cccc}
& \textbf{IP Acc. / F1} & \textbf{OOP Acc. / F1} & \textbf{IP Succ. / Flips} & \textbf{OOP Succ. / Flips} \\ \midrule
\textbf{Known} & & & & \\
 Predicting $y$& 76.3 / 80.3 & 51.5 / 67.2 & --/-- & --/-- \\
Avg. per $c \in \bar{C}$& 89.5 / 94.0 & 45.3 / 46.5 & --/-- & --/-- \\ \midrule
\textbf{Unknown}& & & & \\ 
$\phi = 0.7$& 72.4 / 76.6 & 49.3 / 57.8 & 87.8 / 9. 3& 1.4 / 70.3\\
$\phi = 0.3$& 74.4 / 78.1 & 49.1 / 39.1 & 86.8 / 4.9 & 1.2 / 33.6 \\
$\phi = 0.1$& 76.9 / 79.8 & 50.7 / 16.4 & 86.2 / 1.8 & 1.2 / 9.1 \\
        \bottomrule
    \end{tabular}
    \label{tab:frisian}
\end{table*}

\section{Discussion}\label{sec:discussion}

\subsection{Theory}

\thmref{ndacorrectness} states that when the labeller has high accuracy, a low $\phi$ will be able to reproduce the label set. 
It is then possible to measure the successes to establish a confidence bound on the aggregate labels provided by the evaluator. 
It, however, has a different interpretation: when the accuracy of the evaluator is zero, there is no way to determine the labelling, regardless of $\phi$--it is completely unknowable. 
We argue that this makes sense: otherwise, it would be possible to create information out of thin air, and without any prior knowledge of $f$. 
Hence the primary purpose of the No-Data Algorithm is to establish trustworthiness, not to label data.

Our requirement that $Y$ is binary is sufficient from a theoretical angle, but worth discussing further. 
Arities larger than $2$ may be addressed with class-based separation, at the expense of a linear-time increase in runtime. 
This is an upper bound, and we conjecture that there exist versions of the No-Data Algorithm with smaller overheads. 
We expand upon this and other extensions to the theory in \appref{extensions}. 
There we also show that, experimentally, our algorithm retains its guarantees of correctness under this setup. 
We leave its theoretical characterisation for further work.

\subsection{Experiments}
In both experiments both the LLM and DT evaluators consistently had accuracies close to the known cases in IP/OOP, but distinct success rates in IP and OOP. 
This indicates that the No-Data Algorithm is robust to evaluator strategies attempting deception. 

The large number of flips in the $\phi = 0.5$ DT OOP case (\secref{dt}) and $\phi = 0.3$ LLM OOP (\secref{lowres}) support our interpretation of \thmref{ndacorrectness}. 
Since the model's accuracy is near-random in the OOP dataset, the verifier will almost always catch a lie. %
Hence, most labels were flipped and no information was gained \textit{from the labelling}. 
It is worth noting that \textbf{the success rate remained steady regardless of the value of $\phi$}, even when the flips and accuracies didn't. 
Thus, success rate is the main metric to measure: $\phi$ may be tuneable \textit{post-hoc} to attain better performance. 

In our synthetic data scenario (\secref{dt}), we noted that LLMs struggled with generating data for that problem, and picking it from a set was more effective. However, this was not the case in the natural-language scenario (\secref{lowres}). 
We argue that this is because LLMs are primarily trained with natural language, and identification (classification) and generation are two distinct problems \citep{Kleinberg}; and certain automata-based problems like ours are harder for LLMs \citep{iclnoticl}. 
From the perspective of our work, however, \textit{how} an evaluator generates a datapoint makes no difference, so long as it can prove that it knows what it is doing. 

Finally, remark that the evaluation of LLMs with LLMs, as in \secref{lowres}, depends strongly on the LLMs themselves. 
This is not a circular argument: the verifier (i.e., the prompt) was tuned on the IP case, and asked to evaluate an unknown LLM (r. evaluation prompt) in the OOP case, as done in \secref{lowres} and \appref{linguisticcompetency}, \textit{even when it had near-random performance in this task}. 
This is because the evaluator and verifier are solving different tasks, and hence why the accuracies adjusted to the known cases. 
We provide experiments in \appref{fullresults}--and specifically, \appref{linguisticcompetency}--to confirm this with different LLMs. 
However, from the perspective of future research directions, it raises the question on \textit{what} are the (literature's) results measuring. 
We come back to this point in the next section.

\subsection{The Need for a Rubric}
A key limitation of the No-Data Algorithm is the requirement of a well-designed rubric, and its performance hinges on it. 
If the rubric or aggregator is poorly defined, confidence in the algorithm's results may be misplaced, and its reliability bounded by rubric quality. 
In other words, the No-Data Algorithm measures what it is told to measure. 

Having a rubric is a fundamental aspect of the scientific method: verifying a hypothesis $H_0$ requires measuring a set of experimental outcomes, which are then used to accept or reject $H_0$. 
For this, outcomes must be interpreted (i.e., evaluated as criteria) and then aggregated. 
Thus, the responsibility for designing an effective rubric depends on the scientist (user), not the algorithm. 
This requirement is not unique to our work, or even annotation in machine learning. 
For example, there is work on rubric-based learning in reinforcement learning \cite{he2025advancedifrubricbasedbenchmarkingreinforcement,shao2025drtulureinforcementlearning,huang2025reinforcementlearningrubricanchors,masters2025arcanemultiagentframeworkinterpretable}; and in evaluation \textit{with} (not of) LLMs-as-judges \cite{lei2025dacompbenchmarkingdataagents}. 

However, needing a rubric and characterising it are two distinct considerations. 
For the latter, there are two cases that merit special attention: decomposability, and ambiguity. 
For decomposability, note that the algorithm requires this property for the proofs, but not necessarily at runtime. For example, the experiments from \secref{lowres} used a decomposable rubric only in the OOP case. 
Since the No-Data Algorithm is agnostic to the data labels ($\sigma$ is unknown), it worked as intended. 

Regarding ambiguity, recall that the rubrics from \appref{rubrics} were ambiguous, and the experiments nevertheless succeeded. 
This further illustrates our argument on $H_0$: the performance of the LLMs on the (human-annotated) data was measured with respect to the human labels, which themselves were based on the rubric. 
Both annotators and LLMs used the same rubric. 
Hence, the measurement performed in \secref{lowres} is whether the evaluator could act as a labeller the same way a human would.

\section{Conclusion}\label{sec:conclusion}

Trust in a measurement, although foundational to the scientific method, is difficult to determine when labelled data is scarce. 
In this work we introduced the No-Data Algorithm to formally establish trust in an evaluator when labelled data does not exist. 
This algorithm works by posing carefully-designed challenges, inspired by zero-knowledge proofs, to an evaluator. 
We proved mathematically and empirically that this method is sufficient to establish an evaluator's trustworthiness, or, alternatively, to detect deception by either sycophantic or uninformed models.

Our experiments in synthetic and natural-language scenarios showed close agreement to the theory, as well as some robustness to realistic settings (e.g., ambiguous rubrics and undefined criteria). Ablation studies with other LLMs and label arities showed extensibility and reinforced the findings. 

In our theoretical work we noted that an implication of the No-Data Algorithm is that, when the evaluator cannot be fully trusted, the data cannot be annotated. 
This is because there is no prior information on the data, and hence it would be effectively creating information out of thin air. 
This is why the algorithm's focus is on the success rate. 
If an evaluator has low accuracy, but high success rate, it is very likely uninformed (e.g., the IP case in the experiments). High accuracy but low success rate indicates deception (r. OOP). 
Hence, 
it is up to the user to decide (to their own margin of trust) whether to accept or reject the labelling with respect to the rubric they designed. 
As noted in \secref{ethics}, we urge caution when interpreting the algorithm's outputs, as they depend on the rubric itself.

We argue that our work reframes how evaluator reliability can be measured in situations with scarcely-labelled data. By establishing formal measures of trust, it is now possible to perform rigorous evaluations of `hard' domains such as low-resource languages and medicine. 
This is particularly important due to the debates on the reliability and deployment-readiness of LLMs-as-judges, especially in sensitive applications. 

In terms of further work, there are two open questions remaining, not addressed here due to scope: (1) how can one fully evaluate a model's (not a prompt's) trustworthiness? This is particularly important for open-ended scenarios, such as dialogue. 
Here, the principles behind the EV protocol still apply, albeit the probabilistic guarantees and challenges would differ. 
Hence, (2) what practical adaptations are needed for other data types and output signatures? 
Mathematically the full problem is solvable by assuming a binary string, but empirically this is often not feasible. 

Our work shows that it is possible to judge an evaluator's--such as an LLM-as-a-judge's-- trustworthiness with formal guarantees of correctness. 
It also enables practitioners to transition from assumed evaluator competence, to provable trust, thus enabling the building of more robust and trustworthy AI systems.

\section{Impact Statement}\label{sec:ethics}

This work's contributions involve the ability to establish trust in an LLM. 
A possible risk is over-reliance on this algorithm's measures of trust, especially if the underlying rubric is ill-posed. 
Users should understand that the No-Data Algorithm is only as valid as the rubrics provided. 
We thus urge users to not correlate high success rates with LLM infallibility. 
For the data annotation, all annotators were contracted through an annotator services company, and compensated based on seniority, at a rate starting at 28.7 USD/hr. 
Annotators were encouraged to take breaks and emphasised quality over speed through the annotation instructions.

\bibliographystyle{icml2026}
\DeclareRobustCommand{\DE}[3]{#2}
\DeclareRobustCommand{\VAN}[3]{#2}
\bibliography{biblio}

\newpage
\appendix
\onecolumn
\appendix

\section{Detailed Proofs}\label{app:proofs}

In this section we provide more careful treatments of the proofs of \lemref{pvpcorrectness} (\appref{pvpcorrectnessfull}) and \thmref{ndacorrectness} (\appref{ndacorrectnessfull}). 

\subsection{Correctness of the EV protocol}\label{app:pvpcorrectnessfull}

The key in this proof is to note that accepting a label does not mean it will be the correct label, \textit{but} that the challenges in the EV protocol are sufficient to ascertain said correctness with some probability. 

We separate our proof in two: the first part concerns itself with \textbf{completeness}, or whether the challenges will provide sufficient information to determine if the evaluator can be lying. 
The second part deals with \textbf{robustness} under lying, cheating, or lucky scenarios by the evaluator.

\subsubsection{Part 1 (Completeness)}
We prove the following lemma:
\begin{lemma}\label{lem:evcorrectnessminilemma}
Let $x, x' \in X$ be two binary strings such that $x \neq x'$. 
To determine if the evaluator is lying, the checks from Challenge 1 and Challenge 2 are necessary and sufficient. 
In other words, it is necessary and sufficient to check if $x$ is similar to $x'$.
\end{lemma}
\begin{proof}

(Necessary) It follows from the definition of $C$. Since $f(x) = f(x')$, the check for $x \equiv x'$ is a necessary condition, but not sufficient. 
This challenge yields information around the internal structure of $x, x'$ up to permutation. 
It is necessary because it is possible to have two $x, x'$ such that $x' \not\in \{\pi_i(x)\}$ (i.e., two strings that aren't permutations of one another), and yet $C(x') = C(x)$. However, this does not prove that $x$ is similar to $x'$, since there could be a subset $t \in S_{x'}. t \not\in S_{x}$ and $c(t) = c(s)$ for some $c \in \C$ (for example, if $c$ contains a disjunctive clause). 
Hence running this challenge provides information as to whether the encoding of $C(x')$ contains all information fed into $\sigma$ for a specific $x, x'$. 
It accounts for variation in internal structure of the string so that $C(x) = C(x')$, but does not account for the ordering of these bitstrings. 

On the other hand, $x \equiv x'$ will determine whether the encoding for the bitstrings lead to the same valuation, by checking strict equality. 
This is because, under our setup, the definition of similarity for $x, x'$ implies $\sigma C (x') = \sigma C (x)$. 
This makes it order-sensitive, but does not account for the internal structure of the string. 

Since neither of them provide information about the other, it follows that both are needed to fully determine $f(x) = f(x')$. 

(Sufficient) It follows directly from the above. 
Knowing $C(x) = C(x')$ solely implies that the final encodings for $x$ and $x'$ are equivalent, not that they are structurally equivalent. 
Conversely, knowing $\bar{C}(x), \bar{C}(x)$ does not necessarily yield information as to \textit{how} these are converted into $C(x)$ and $C(x')$. 
Hence it is sufficient to know both quantities to determine if $x$ is similar to $x'$. 

Consequentially, testing if $x$ is similar to $x'$ naturally implies that these challenges are sufficient to test $f(x) = f(x')$ (via $x \equiv x'$), and necessary (via $x \cong x'$) to determine if the evaluator is lying. 

This concludes the proof. 
\end{proof}

\subsubsection{Part 2 (Robustness)}
We prove the following lemma:
\begin{lemma}\label{lem:evrobustnessminilemma}
The probability of error (i.e., the verifier accepts a $y' \neq y$) at the $r^{\text{th}}$ round is $\left(1/4\right)^r$. 
\end{lemma}
\begin{proof}
In here we have two scenarios: either the evaluator knows $f$, or it doesn't, and hence it is lying or cheating. 

Knowing $f$ implies that the evaluator will generate a `good' $x'$ every time, such that $\tilde{y'} = y$; 
and hence it will pass the checks from the verifier w.p.1. 
For the proof we focus then on the second case: that the evaluator is lying or cheating. 

There are then two (disjoint) possibilities for a generated $\langle x', y' \rangle$: either that the evaluator is lying, but gets lucky; or that it is cheating and can generate the answer for one of the challenges. 
Remark that generating an answer to pass \textit{both} challenges is equivalent to knowing $f$. 
We thus focus on the two possible lies from the evaluator, which would allow it to pass either challenge.

\subsection{Lie 1: Generate a `good' $x'$}
Suppose that the evaluator generates an $x'$ such that $S_{x'} = S_{x}$, and thus can pass Challenge 1. 
It cannot pass Challenge 2, since $S_{x'} = S_{x}$ implies that either $C(x') \neq C(x)$ (and hence it fails it); or $C(x') = C(x)$ (and hence either it gets lucky, or it knows $f$). 
Since the challenges appear with probability $1/2$, the likelihood that the evaluator gets lucky in this round is $1/4$. 

\subsection{Lie 2: $f(x') \neq y'$}
Now suppose that the evaluator anticipates Challenge 2, and picks an $x'$ such that $C(x') = C(x)$. 
Similar to Lie 1, then either $S_{x'} \neq S_{x}$ (and thus it fails Challenge 1); or $S_{x} = S_{x'}$ (and hence it gets lucky or it knows $f$). 
The likelihood of getting lucky this round is again $1/4$. 

Therefore, the probability of the verifier to \textit{not} catch the evaluator's lie is $1/4$ in any round. At the $r^{\text{th}}$ round, $\left(1/4\right)^r$. 
This concludes the proof. 
\end{proof}

It follows from \lemref{evcorrectnessminilemma} that both challenges are necessary and sufficient to decide if an evaluator is lying; and from \lemref{evrobustnessminilemma} that the ability of the verifier to do this is bounded by a probability of error of $(1/4)^r$. 
Putting both lemmas together we obtain the proof for \lemref{pvpcorrectness}.

\subsection{Correctness Bounds of the No-Data Algorithm}\label{app:ndacorrectnessfull}

By a simple application of the union bound on the probability that the evaluator was wrong, but not detected by the verifier, and the probability that the evaluator was wrong, detected by the verifier, but without having its label flipped. 

For this, it is easier to work with errors $\epsilon = 1 - \alpha$, and remark that, from \lemref{pvpcorrectness} the first term (undetected, no flip) is given by 

\begin{align}
\Pr[\text{undetected}] &= \sum_{f(x) \neq y}\left(\frac{1}{4}\right)^r\text{, and hence} \\
\mathrm{E}[\text{undetected}] &= \epsilon\left(\frac{1}{4}\right)^r.
\end{align}

Likewise, the second term accounts \textit{only} for the situation where the mislabelling was detected and the label was not flipped:
\begin{equation}
\mathrm{E}[\text{detected, no flip}] = \epsilon\left(1 - \phi\right)\left(1 - \left(\frac{1}{4}\right)^r\right) .
\end{equation}

This yields:
\begin{align}
\bigcup \Pr[\text{wrong}] &\leq \Pr[\text{undetected}] + \Pr[\text{detected, no flip}]\\    
&\leq \left(\frac{1}{4}\right)^r +\left(1 - \phi\right)\left(1 - \left(\frac{1}{4}\right)^r\right)\\
\mathrm{E}[\text{wrong}] &\leq \epsilon\left(1 - \phi + \phi\left(\frac{1}{4}\right)^r\right)\text{, by linearity of expectation.}
\end{align}

Substituting back $\Pr[\text{wrong}] = 1 - \Pr[\text{right}]$ and $\epsilon = 1 - \alpha$ yields the desired value. This concludes the proof.

\section{Characterisation Bounds}\label{app:bounds}

\begin{lemma}
    Suppose the evaluator and verifier run in polynomial time for any input $x \in X$. 
    Then, if the No-Data Algorithm is ran with a dataset  of size $\size{D}$ and with $r$ rounds of EV protocol, the runtime is then \BigO{r\size{D}}.
\end{lemma}
\begin{proof}
    Straightforward by noting the number of calls in the algorithm.
\end{proof}

\begin{remark}
The runtime of the No-Data Algorithm is linear (up to a factor of $r$), but this algorithm is designed to calibrate an evaluator, \textit{not} label a full dataset. 
Thus, $D$ and $r$ may be small, as in \secref{lowres}. 
\end{remark}

\section{Extensions}\label{app:extensions}

\subsection{$k$-ary Label Sets}\label{app:kary}
To convert a dataset of arity $k>2$, it is simply a matter of following a `one-vs-all' breakdown, rerunning the No-Data algorithm $k - 1$ times. 
On each run, the label set $\{l_1, \dots, l_k\}$ is mapped as $\{l_1\}, \{l_2, \dots, l_k\} \mapsto \{0\}, \{1\}$. In these cases, the runtime of the algorithm is increased by a factor of $k - 1$, but the theoretical guarantees hold. 

It is then worthwhile considering whether it is possible to circumvent the runtime by simply running the No-Data algorithm without any modifications, except the label flip, and still obtain similar results. 
To test this, we ran the same setup from \secref{lowres}, but splitting the aggregator into three labels:
\begin{enumerate}
    \item Label 0: All of the $c_i \in \bar{C}$ evaluate to zero.
    \item Label 1: All, except one, of the $c_i \in \bar{C}$ evaluate to zero.
    \item Label 2: None of the $c_i \in \bar{C}$ are zero.
\end{enumerate}

The results are in \tabref{llmklabel}. 
Overall, it is possible to see that the results largely adhere to the previously-observed binary label settings. Namely, lying evaluators have near-zero successes and their original performance is close to the known case, with a best-of difference of 6\% accuracy. 
This difference is comparable to the binary case in \tabref{frisianfull} for an untuned $\phi$. 
Indeed can observe a slight improvement with respect to the known case when the label flip is disabled. As shown in \secref{correctness} (and experimentally in \appref{generator}), the label flip is required to achieve full restoration of the original score, which will require further algorithmic developments for a $k$-ary setting, and are beyond the scope of this work. 
In sum, while empirically the No-Data Algorithm may be used for $k>2$, further theoretical development is needed to extend this algorithm to non-`one-vs-all' strategies with the same guarantees. 

We then may pose the following:
\paragraph{Conjecture 1:} Let $Y$ be a label set with arity $k > 2$, and $D$ a dataset. 
Then, for any number of rounds $r$, there exists an $n$ such that a version of the No-Data Algorithm runs in $\text{O}(nr\vert D\vert)$ steps, where $1 < m < n < k - 1$ and $m$ depends on $k$. 

For a `version' of the No-Data Algorithm, we mean that it maintains the same correctness bounds from \thmref{ndacorrectness}. 

\begin{table}[]
     \centering
    \caption{Ablation study reframing the setup from \secref{lowres} into a ternary-labelled experiment (${0, 1, 2}$). 
    We report the results with accuracy and $F_\mu$ (macro) score. 
    For comparison, a random guesser obtains 33 accuracy and 30 $F_\mu$. 
    In this experiment, the known case was still reconstructable from the algorithm's run, although with more noise. Disabling label flips improved restoration of the original scores. 
    Full reconstruction would require the development of further algorithmic strategies to retain all guarantees from the No-Data Algorithm. 
    }
    \begin{tabular}{l|cccc}
& \textbf{IP Acc. / F$_\mu$} & \textbf{OOP Acc. / F$_\mu$} & \textbf{IP Succ. / Flips} & \textbf{OOP Succ. / Flips} \\ \midrule
\textbf{Known} & & & & \\
 Predicting $y$& 71.8 / 61.1 & 10.5 / 7.0 & --/-- & --/-- \\ \midrule
\textbf{Unknown}& & & & \\ %
$\phi=0.1$ & & & & \\ %
\quad Flip    & 66.0 / 55.7 & 12.2 / 10.8 & 39.4 / 4.3 & 0.4 / 11.3  \\
\quad No flip & 67.0 / 57.1 & 10.1 / 6.4 & 39.4 / -- & 0.4 / -- \\
$\phi=0.3$ & & & & \\ %
\quad Flip    & 59.8 / 52.3 & 20.0 / 17.8 & 38.1 / 18.1 & 0.8 / 34.2 \\
\quad No flip & 61.6 / 55.1 & 10.5 / 7.0 & 38.1 / -- & 0.8 / -- \\
$\phi=0.7$ & & & & \\ %
\quad Flip    & 39.5 / 40.0 & 30.9 / 21.1 & 35.3 / 47.4 & 0.8 / 71.6 \\
\quad No flip & 46.8 / 43.4  & 0.10 / 6.7 & 35.3 / -- & 0.8 / -- \\

    \end{tabular}
    \label{tab:llmklabel}
\end{table}

\subsection{Distance-based Constraints}
The constraint that $C(x) = C(x')$ is too strict for some applications. We sketch out in this section a variant of the No-Data Algorithm where, instead of requiring equality, we request closeness with respect to some arbitrary metric $\delta$ in Challenge 2. 
Suppose now that the failure is given whenever $\delta(C(x), C(x')) \geq \epsilon$, for some $\epsilon \in [0, 1/2]$. 
The bounds of correctness are roughly equivalent to these described in \lemref{pvpcorrectness} and \thmref{ndacorrectness}. However, the term $(1/4)^r$ now takes a dependency on the probability of generating a string that is far (based on $\delta$) from the desired constraint, and thus the bounds become weaker when said probability is larger than $1/2$.

\section{Full Results}\label{app:fullresults}

\subsection{LLM-Based Results}

In a full version of our study from \secref{dt}, we evaluated multiple open and closed LLMs. 
The LLMs studied were o3-mini (as reported in the main section), DeepSeek \citep{deepseekai2025deepseekr1incentivizingreasoningcapability}, GPT-4o, and Qwen 2.5 VL 7B \citep{qwen2.5-VL}, all under the same experimental setup from \secref{dt}. The results can be found in \tabref{llmablationcomparison}. 
Outside of o3-mini, all models struggled with this problem, even in the known case. Consequentially, their performance in the unknown case also adjusted to the predictions. 
It is worth noting that, even when their accuracy was (relatively) high in OOP, their success rate was low--as expected. This is particularly noticeable in o3-mini, GPT-4o, and DeepSeek. 

\begin{table}[ht]
    \centering
    \caption{Comparison of the No-Data algorithm with respect to various LLMs and values of $\phi$. 
    The known case reflects the LLM's expected performance in the dataset, while the unknown case shows what the No-Data Algorithm outputs. 
    All LLMs broadly adjusted to their known case. 
    The LLMs also showed on average low accuracies across the board. 
    However, success rates were markedly different between the known and unknown cases under high accuracy, indicating possible deception. 
    When the accuracy was low, success rates remained low regardless of the setting. 
    }
    \begin{tabular}{lcccc}
          \toprule
  & \textbf{o3-mini} & \textbf{DeepSeek} & \textbf{GPT-4o} & \textbf{Qwen} \\
\midrule
\textbf{Known Case} & & & & \\
\midrule
\textbf{IP} & & & & \\
Accuracy/$F_1$ & 99.8 / 99.8 & 61.0 / 70.8 & 60.1 / 69.7 & 50.0 / 66.7 \\
Successes/Flips & --/-- & --/-- & --/-- & --/-- \\
\textbf{OOP}&  & & & \\
Accuracy/$F_1$& 59.0 / 70.0 & 54.4 / 65.6 & 55.8 / 66.4 & 50.2 / 66.8 \\
Successes/Flips& --/-- &--/-- & --/-- & --/--\\
\midrule
\textbf{Unknown Case} & & & & \\
\midrule
(\textbf{$\phi = 0$}) & & & & \\
\textbf{IP} & & & &\\
Accuracy/$F_1$  & 54.0 / 58.9 & 46.8 / 18.0 & 42.8 / 23.6 & 54.1 / 27.3\\
Successes/Flips & 53.8 / 46.2 & 12.9 / 87.1 & 6.6 / 93.4 & 13.3 / 87 \\
\textbf{OOP} & & & & \\        
Accuracy/$F_1$  & 58.4 / 56.6 & 55.0 /39.5 & 78.7 / 78.1 & 33.5 / 0.0 \\
Successes/Flips & 26.3 / 73.7 & 16.5 / 39.5& 34.1 / 65.9 & 16.3 / 83.7 \\
(\textbf{$\phi = 0.5$}) & & & & \\
\textbf{IP} & & & & \\
Accuracy/$F_1$  & 81.6 / 82.4 & 53.2 / 52.4 & 48.6 / 48.8 & 53.2 / 54.2 \\
Successes/Flips & 55.4 / 5.0 & 12.5 / 44.2 & 6.6 / 47.8 & 13.1 / 47.8 \\
\textbf{OOP} & & & & \\      
Accuracy/$F_1$  & 56.4 / 59.3 & 55.4 / 58.0 & 63.3 / 67.4 & 44.4 / 54.2 \\
Successes/Flips & 28.7 / 59.3 & 19.5 / 37.8 & 33.7 / 34.1& 16.3 / 47.8 \\
(\textbf{$\phi = 0.9$}) & & & & \\
\textbf{IP} & & & & \\
Accuracy/$F_1$  & 94.2 / 94.2 & 60.0 /68.4 & 56.6 / 63.8 & 51.2 / 65.4 \\
Successes/Flips & 55.4 / 5.0 & 11.9 / 9.4  & 5.8 / 10.8 & 13.1 / 8.8\\
\textbf{OOP} & & & & \\        
Accuracy/$F_1$  & 60.6 / 65.5 & 53.2 / 63.9 & 53.4 / 63.6 & 50.2 / 64.9 \\
Successes/Flips & 28.7 / 7.8 & 18.1 / 7.6 & 34.5 / 4.6 & 16.3 / 8.0 \\
\bottomrule
    \end{tabular}
    \label{tab:llmablationcomparison}
\end{table}

\subsection{Ablation: Lying Evaluators and Lying Generators}\label{app:lying}

In this study we evaluated the impact of various aspects of lying in evaluation, both during generation and evaluation (of $C$). The results are in \tabref{lies}. 
The lies tested were when the model claims to:
\begin{enumerate}
    \item know $\sigma$: that is, it can output an encoding $C(x)$, but does not necessarily know $\bar{C}(x)$; 
    \item know $f$: that is, it can retrieve a datapoint $x' \in X$ such that $y' = y$, but nothing else; and
    \item approximately know $f$ up to a probability $p = 1/10$ (r. it could pass both challenges with probability $1 - p$). 
\end{enumerate}
The last `lie' is more akin to how a realistic generator would behave (e.g., an LLM), sometimes understanding the challenges well, but failing them with some probability. 

As before, the results indicate that lying evaluators tend to score low in the success rate, even when their accuracies are close to the known case. 
Remark that (a) the evaluator that lies on knowing $\sigma$ had a success rate much higher than that of lying of knowing $f$; and (b) the evaluator with approximately knowledge of $f$ had a comparatively low (44\%) success rate. 

This success rate depends on the cardinality of $\bar{C}$: when it is equal to $\C$, the model will always pass Challenge 1 (and fail Challenge 2). 
However, when $\size{\bar{C}} > \size{C}$, there will be criteria that the evaluator cannot possibly solve without knowing how $C(x)$ is constructed based on its internal structure, and hence it will fail them with some probability.

\begin{table}[]
    \centering
    \caption{Performance for a lying DT evaluator on the No-Data Algorithm with $\phi = 0.4$. 
    From top to bottom: Lie 1: $\sigma$ unknown. Lie 2: $f$ unknown. Lie 3: $f$ approximately known (lying with probability $p=1/10$) in the generation step.
    In all these calls, the evaluator itself also outputs the incorrect label with probability $p=1/10$. 
    Note the comparatively high success rates in the IP case of Lies 1 and 2. We discuss these findings in \secref{discussion}. 
    Also note the consistently low success rate for OOP across all experiments, thus indicating the robustness of this algorithm to any evaluator strategy in scenarios where the data is completely unknown to the evaluator. 
    }
    \begin{tabular}{l|cc|cc}
    &  \textbf{IP (Flips)} & \textbf{IP (No Flips)} & \textbf{OOP (Flips)} & \textbf{OOP (No Flips)}  \\
    \toprule 
     \textbf{No Lie}: DT & & & & \\
       Successes / Flips        & 100.0 / 0.0 & 100.0 / ---- &  6.0 / 35.3 & 6.4 / ---- \\ 
       Accuracy/ F$_1$          & 60.6 / 58.8 & 59.0 / 57.4 & 48.6 / 47.3 & 45.6 / 44.8 \\
    \midrule
     \textbf{Lie 1}: $\sigma$ unknown  & & & & \\
       Successes / Flips        & 17.0 / 35.1 & 15.1 / ---- & 2.8 / 35.1 & 2.2 / ---- \\ 
       Accuracy/ F$_1$          & 58.6 / 58.8 & 42.6 / 45.0 & 52.4 / 53.8 & 49.0 / 49.4 \\
    \midrule
     \textbf{Lie 2}: $f$ unknown  & & & & \\
      Successes / Flips        & 0.6 / 37.4  & 0.2 / ---- & 3.0 / 35.7  &  2.8 / ---- \\ 
      Accuracy/ F$_1$          & 50.6 / 48.1 & 39.4 / 42.4 & 51.4 / 52.2 & 51.2 / 52.8 \\ 
    \midrule
     \textbf{Lie 3}: generator w.p. $p=1/10$  & & & & \\
      Successes / Flips        & 43.6 / 24.9 & 51.3 / ----  & 4.4 / 37.9  & 4.6 / ---- \\ 
      Accuracy/ F$_1$          & 55.8 / 54.2 & 49.0 / 49.4  & 47.6 / 49.5 & 46.6 / 48.6 \\ 
    \bottomrule
    \end{tabular}
    \label{tab:lies}
\end{table}

\subsection{Ablation: LLM Generation Strategies}\label{app:generator}

In this study we accounted for the sensitivity to the prompt. This is because it is likely that an LLM's performance in the EV protocol could be marked by \textit{how} it generates the datapoints. 
We tested two different generation strategies: picking a matching datapoint from the data, as in \secref{results}; and actually generating a datapoint from the rubric. 
The dataset is the synthetic dataset, and we compared the performance of the LLM from \secref{dt} (o3-mini); and GPT-4o \citep{gpt4o}. The results are in \tabref{llmablation}. 
We found that, again, the experimental results adjusted to the predictions, with lying and low-performing models being easily spotted by the No-Data Algorithm. 
However, we also found that actual generation (versus picking) led to lower performance in both LLMs; with o3-mini having a slightly lower drop (-3\%) than GPT-4o (-5\%). 

\begin{table}[]
    \centering
    \caption{Ablation study on LLM generation strategies, comparing a high-performing model (o3-mini) with a lower-performing model (GPT-4o). 
    We report the results at $\phi = 0.1$. 
    Both LLMs were able to broadly adjust to their original accuracy, but success rates are different when they are generating a datapoint versus picking it. 
    }
    \begin{tabular}{l|cc|cc}
          & \textbf{o3-mini (IP)} & \textbf{o3-mini (OOP)} & \textbf{GPT-4o (IP)} & \textbf{GPT-4o (OOP)} \\
          \toprule
        \textbf{Picking} & & & & \\
        Accuracy/F$_1$ & 97.6 / 97.6 & 59.0 / 70.0 & 58.0 / 67.1 & 56.4 / 65.4 \\ 
        Successes/Flips& 81.3 / 1.8 & 27.9 / 6.0 & 28.7 / 71.3 & 10.8 / 8.8 \\
        \midrule
        \textbf{Generating} & & & & \\
        Accuracy/F$_1$ & 94.2 / 94.2 & 60.6 / 65.5 & 56.6 / 63.8 & 53.4 / 63.6 \\
        Successes/Flips& 55.4 / 5.0 & 28.7 / 7.8 & 5.8 / 10.8 & 34.5 / 4.6 \\
        \bottomrule
    \end{tabular}
    \label{tab:llmablation}
\end{table}

\subsection{Ablation: Linguistic Competency}\label{app:linguisticcompetency}

In this section we present an expanded version of the results from \secref{lowres}, including the performance of Qwen 2.5B VL 7B on the same corpus, with GPT-4.1 as a verifier. 
This model has no indications of being proficient in West Frisian. On the other hand, the calibration of GPT-4.1 in IP through the known case indicates that this verifier is capable of acting as a comparatively reliable component. 
Indeed, in the IP case, where the model was given the correct rubric (i.e., it did not lie), the success rate for the Qwen-based evaluator was low--thus indicating inability to solve the problem, as opposed to deception. 
Consequentially, in the OOP case (where it was set up to lie), the success rate was lowest. 
As before, this number remained steady throughout the experimentation. 
The full results are in \tabref{frisianfull}. 

\begin{table}[]
    \centering
    \caption{Performance of Qwen 2.5B and GPT-4.1 on the natural-language version of the No-Data Algorithm with our ablation study. 
    In this setup, Qwen 2.5B and GPT-4.1 act as evaluators, and GPT-4.1 as a verifier. 
    The success rate for Qwen 2.5B was low in both the IP case (implying that the the model is not lying; just wrong) and the OOP case (where Qwen 2.5B was lying \textit{and} wrong). Since the success rate is much lower in the latter, it follows that the No-Data Algorithm is particularly proficient at detecting confident-but-wrong evaluators. 
    }
    \begin{tabular}{l|cccc}
& \textbf{IP Acc. / F$_1$} & \textbf{OOP Acc. / F$_1$} & \textbf{IP Succ. / Flips} & \textbf{OOP Succ. / Flips} \\ \midrule
\textbf{Known} & & & & \\
 Predicting $y$& 76.3 / 80.3 & 51.5 / 67.2 & --/-- & --/-- \\
Avg. per $c \in \bar{C}$& 89.5 / 94.0 & 45.3 / 46.5 & --/-- & --/-- \\ \midrule
\textbf{Unknown}& & & & \\ \midrule
GPT-4.1 & & & & \\ \midrule
$\phi = 0.7$& 72.4 / 76.6 & 49.3 / 57.8 & 87.8 / 9. 3& 1.4 / 70.3\\
$\phi = 0.3$& 74.4 / 78.1 & 49.1 / 39.1 & 86.8 / 4.9 & 1.2 / 33.6 \\
$\phi = 0.1$& 76.9 / 79.8 & 50.7 / 16.4 & 86.2 / 1.8 & 1.2 / 9.1 \\\midrule
Qwen 2.5B & & & & \\ \midrule
$\phi = 0.7$& 62.5 / 63.9 & 51.7 / 59.5 & 34.4 / 46.0 & 1.8 / 70.3\\
$\phi = 0.3$& 55.9 / 64.8 & 51.5 / 43.2 & 34.2 / 21.4 & 2.1 / 34.4 \\
$\phi = 0.1$& 55.5 / 67.7 & 49.7 / 19.8 & 35.0 / 6.2 & 2.1 / 11.5 \\
        \bottomrule
    \end{tabular}
    \label{tab:frisianfull}
\end{table}

\subsection{Comparison With Baselines}\label{app:baseline}

It was noted in \secref{experiments} that baselines for this problem remain challenging, as they are unable to spot lying evaluators. 
Ensemble methods, for example, provide `a' label, but not a measure of trust. Typically, this would be given by agreement metrics such as pairwise Cohen's $\kappa$ or percentage agreement (PA). 
However, these metrics must be interpreted and calibrated, which in turn depends on the problem at hand \cite{10.1145/3485447.3512242}. 
Likewise, they have other drawbacks, such as sensitivity to class imbalance \cite{CICCHETTI1990551,gwet}. 
These issues are hard to circumvent when there do not exist labels or references for calibration. 
On the other hand, in this scenario the No-Data Algorithm matches the known-case performance, provides a trust metric (the success rate), and spots deception.

In this section we illustrate this by comparing two consensus-based strategies (ensembles): aggregating the predictions of three evaluators (GPT-4.1, o3-mini, and Qwen-2.5) by averaging them, and by taking a majority vote. 
The comparisons are with respect to the known case from \secref{lowres}, for GPT-4.1 and including IP and OOP. 
The agreement metrics are Cohen's $\kappa$ and PA. 
The full results are in \tabref{ensemble}. 

Perhaps unsurprisingly, in terms of baselines, ensembling three evaluators matched or outperformed the accuracy and F$_1$ for the (single model) IP known case. 
Nonetheless, they were marked as presenting poor agreement ($\kappa = 0.38$; PA = $0.72$), as per \citet{fleiss}. 
Without calibration, poor agreement would indicate caution. Indeed, this value should be on average higher since the classes are balanced and the ensemble performances are good \cite{10.1093/ptj/85.3.257}. 

Moreover, when in the presence of lying evaluators (OOP), ensembling methods had a difficult time. Averaging their predictions correctly predicted random performance (50.1\% accuracy); but majority vote aggregation reported accuracies well above what would be expected (61.6\%), thus indicating vulnerability to deception. 
Even though Cohen's $\kappa$ was reliable in this case (noting only chance agreement, $\kappa = 0.0$), PA reported a moderate agreement ($0.37$). 
In practice, these contradictory findings would require calibrating the metrics, \textit{especially} if the IP case's $\kappa$ is used as justification for the OOP case. 
This is not possible without labels.\footnote{Realistically, in an imbalanced-labels or multi-class scenario, metrics such as PA and Cohen's $\kappa$ are even more sensitive, thus further complicating calibration \cite{CICCHETTI1990551}.} 

In contrast, the No-Data Algorithm provided a way to accept/reject the evaluator's predictions (the success rate; 86.2 for IP and 1.2 for OOP) as well as a slightly higher accuracy (76.9\%) when compared to the single-model IP case. 
Since no calibration is needed for any parameter, we argue that, in the unknown case (the focus of our work), it is much more helpful to use this method over other algorithms. 

\begin{table}[]
    \centering
    \caption{Baseline comparisons between two ensemble methods and the results of the No-Data Algorithm from \tabref{frisianfull}. 
    For measures of trust, we report Cohen's $\kappa$ and PA for the ensembles and success rate for the No-Data Algorithm. 
    In the IP case, Cohen's $\kappa$ marks agreement as poor (as per \citealt{fleiss}); even when the ensemble's performance nears or surpasses the known case. 
    In OOP, the ensembles obtained above-random performance, which is impossible given the setup. 
    Thus, trusting the value of a calibrated $\kappa$ in IP for OOP would yield spurious results. 
    Hence better calibration is needed, since the information provided by the agreement metrics is insufficient (i.e., it is either contradictory or does not indicate full agreement). 
    This is, however, not possible in the unknown case. 
    In contrast, the No-Data Algorithm does not rely on calibrations and is robust to lying evaluators, while also being able to restore the known case accuracy.
    }
    \begin{tabular}{l|cc|cc}
& \textbf{IP trust} & \textbf{OOP trust} & \textbf{IP Acc. / F$_1$} & \textbf{OOP Acc. / F$_1$} \\ \midrule
\textbf{Known} & -- & -- & & \\
GPT-4.1 & & & 76.3 / 80.3 & 51.5 / 67.2 \\
\textbf{No-Data Algorithm}&  86.2 & 1.2 & & \\
GPT-4.1 at $\phi = 0.1$ & & & 76.9 / 79.8 & 50.7 / 16.4 \\
\textbf{Ensembles} & 0.39 / 0.72 & 0.0 / 0.37 & & \\ 
Average       & & & 80.6 / 81.3 & 50.1 / 0.0 \\
Majority vote & & & 75.5 / 79.9 & 61.6 / 48.7
\\
        \bottomrule
    \end{tabular}
    \label{tab:ensemble}
\end{table}

\section{Detailed Experimental Methods}\label{app:experiments}
For our first experiment, we used o3-mini (version \textsc{2025-01-31}) with maximum tokens set to 50,000 and left all other parameters as default. 
GPT-4o (version \textsc{2024-05-13}) and Qwen 2.5 VL 7B were called with maximum tokens set to 1,024 and temperature set to zero. 
DeepSeek (version \textsc{r1-distill-qwen-32b}) also had the temperature set to zero, but the maximum tokens set to 2,048 to account for the longer reasoning trace. 
For the decision tree we used the implementation from sklearn\footnote{\url{https://scikit-learn.org/stable/index.html}} with all parameters, except the random seed, left as default. 

We carried out the second experiment (natural language) with GPT-4.1 (version \textsc{longco-2025-04-14}). The evaluator (labelling, criteria) and verifier were set at 256 maximum tokens, while the generator (i.e., the evaluator generating similar datapoints) was set at 5,000 tokens. All models were called with temperature set to zero. 

LLMs were sometimes prone to return unparseable outputs. To mitigate this, in case of failure, we re-queried the models up to five times. 

All LLMs were called via the Azure OpenAI API. 
All data analysis was performed on a consumer-grade laptop. 
Random seeds are set for the training in the code and indicated in the repository. 

\subsection{Rubrics}\label{app:rubrics}

We present the full rubrics used in our experiments. 
Tables~\ref{rubricifnl} and~\ref{rubricoofnl} have the specifications for the natural-language experiments. Both rubrics were used by the human annotators, and passed verbatim to the LLMs.

\begin{table}[h]
    \centering
    \caption{Rubric for the natural-language evaluation (IP) experiment. This rubric is meant to evaluate the quality of the output of an LLM. The rubric above was given to both the human annotators and the LLM from the No-Data Algorithm.}    \begin{tabular}{lp{0.84\linewidth}}
\textbf{Criterion} & \textbf{Definition} \\\midrule 
c1) & The response must be in West Frisian.\\
c2a) & The response must be culturally (e.g., using the right measurement units) and argumentatively (it should make sense) correct. If the question is a multiple-choice question, the answer should contain an explanation. If it requests code, it should also contain an explanation that is clear. Grammar or accuracy of the response are not measured here.\\
c2b) & The response must be correct. If it is code, it should not have syntax errors.\\
c3) & The response must be grammatically correct: coherent, good spelling, etc. Code syntax is not measured here.\\
c4) & The response must not be cut off.\\
c5) & The model must follow the instructions from the user (the prompt) exactly and completely, even if its answer is wrong. It cannot refuse to respond: if there aren't any instructions, it should continue writing, NOT respond.\\
\bottomrule
    \end{tabular}
    \label{rubricifnl}
\end{table}

\begin{table}[h]
    \centering
    \caption{Rubric for the natural-language evaluation setup (OOP). 
    This rubric was designed to be mostly orthogonal to the IP version. 
    That meant measuring a different problem altogether, albeit with some overlaps (e.g., part of c2b overlaps with c2a in IP, c2a here is c4 in IP). 
    Some criteria are completely different, such as c1 and c5.}    \begin{tabular}{lp{0.84\linewidth}}
\textbf{Criterion} & \textbf{Definition} \\\midrule
c1) & The response must be in Dutch (West Frisian Dutch)\\
c2a)& The response must not be cut off.\\
c2b)& The response must make as many references as possible to Dutch culture.\\
c3) & The response must continue writing if there is no prompt.\\
c4) & (A) The response must not be cut off, or (B) The response must make as many references as possible to Dutch culture. At least one of them must be correct. If they are both zero or one, the response is zero.\\
c5) & The response must provide a summary or conclusion. It must be explicitly marked as `summary' or `conclusion'.\\
\bottomrule
    \end{tabular}
    \label{rubricoofnl}
\end{table}

\subsection{Prompts}\label{app:prompts}
The prompts we used for our LLM evaluation on synthetic experiments are in \tabref{evaluatorprompt} (evaluation) and \tabref{generationprompt} (generation). All exemplars are omitted for brevity. Whenever a parsing error occurred, we retried the call up to five times. If it failed on the fifth, we returned a random string. 
For the natural language experiments the prompts are in \tabref{nlevaluatorprompt} (evaluator/verifier) and \tabref{nlgeneratorprompt} (generator; part of the evaluator). These prompts required no exemplars. When parsing errors occurred, we retried up to five times, and otherwise returned a zero-labelled value.

\begin{table}[h]
    \centering
    \caption{System prompt for the evaluator (synthetic experiment). Anchors (`|reasons|', `|label|') are used for parsing. 
    The text contained within `|reasons|' is, as specified by the exemplars (not pictured) a chain-of-thought approach determining criterion-by-criterion matching. The rubric, explained in natural language, is inserted in the marker. Throughout the experiments we only used the IP rubric.}
    \begin{tabular}{l}
\toprule
\cellcolor{RoyalPurple!20}You are labelling binary strings based on a rubric (given below).\\
\cellcolor{RoyalPurple!20}First return the parts of the criterion that match the string, and the values.\\
\cellcolor{RoyalPurple!20}Then return the label based on the aggregate function.\\
\cellcolor{RoyalPurple!20}\# Rubric: \\
\cellcolor{RoyalPurple!20}NATURAL-LANGUAGE RUBRIC GOES HERE\\
\cellcolor{RoyalPurple!20}\# Aggregation Function:\\
\cellcolor{RoyalPurple!20}Majority vote\\
\cellcolor{RoyalPurple!20}Return your answer in the form:\\
\cellcolor{RoyalPurple!20}$|$reasons$|$\\
\cellcolor{RoyalPurple!20}(list of reasons)\\
\cellcolor{RoyalPurple!20}$|$reasons$|$\\
\cellcolor{RoyalPurple!20}$|$label$|$\\
\cellcolor{RoyalPurple!20}(the label)\\
\cellcolor{RoyalPurple!20}$|$label$|$ \\
\bottomrule
    \end{tabular}
    \label{tab:evaluatorprompt}
\end{table}

\begin{table}[h]
    \centering
    \caption{System prompt for the generator (synthetic experiment). This prompt does not utilise exemplars. The rubric, explained in natural language, is inserted in the marker. 
    Throughout the experiments we only used the IP rubric and selected datapoints from the IP training dataset, ensuring that at least one of the entries matched the criterion. Given that the model did not follow the output format exactly, some trial-and-error was needed to ensure a comparatively high parse success rate.}
    \begin{tabular}{l}
\toprule
\cellcolor{RoyalPurple!20}You are given five datapoints and a set of criteria.\\
\cellcolor{RoyalPurple!20}Based on these, pick the datapoint that matches the criteria exactly.\\ 
\cellcolor{RoyalPurple!20}\# Rubric:\\
\cellcolor{RoyalPurple!20}NATURAL-LANGUAGE RUBRIC GOES HERE\\
\cellcolor{RoyalPurple!20}\# Datapoints:\\
\cellcolor{RoyalPurple!20}DATAPOINTS GO HERE\\
\cellcolor{RoyalPurple!20}Return your rationale, and then your chosen datapoint ONLY in the form:\\
\cellcolor{RoyalPurple!20}$|$datapoint$|$\\
\cellcolor{RoyalPurple!20}(the datapoint)\\
\cellcolor{RoyalPurple!20}$|$datapoint$|$\\
\bottomrule
    \end{tabular}
    \label{tab:generationprompt}
\end{table}

\begin{table}[h]
    \centering
    \caption{System prompt for generation used in the ablation study from \appref{generator}. Unlike \promptref{generationprompt}, this prompt does take exemplars (not pictured): the first five entries in the training dataset, along with chain-of-thought-style rationale.}
    \begin{tabular}{l}
\toprule
\cellcolor{RoyalPurple!20}You are a datapoint generator over binary strings.\\
\cellcolor{RoyalPurple!20}Given a rubric (given below), a datapoint, and a label, return a similar datapoint that has the same\\
\cellcolor{RoyalPurple!20}label, and fulfils the same conditions as the rubric.\\ 
\cellcolor{RoyalPurple!20}For convenience, always start with the same datapoint: 0100. It will be easier to work with.\\
\cellcolor{RoyalPurple!20}\# Rubric:\\
\cellcolor{RoyalPurple!20}NL RUBRIC GOES HERE\\
\cellcolor{RoyalPurple!20}Return your rationale, and then the final datapoint in the form:\\
\cellcolor{RoyalPurple!20}$|$datapoint$|$\\
\cellcolor{RoyalPurple!20}(the datapoint)\\
\cellcolor{RoyalPurple!20}$|$datapoint$|$\\
\bottomrule
    \end{tabular}
    \label{tab:generationablationprompt}
\end{table}

\begin{table}[h]
    \centering
    \caption{System prompt for the natural-language evaluator and verifier. 
    The output format requested is always in JSON. 
    Depending on the setup, the output format and criteria will vary. 
    For example, if it is meant to evaluate a single criterion, say, c2b, the definition will include only the definition of said criterion and the schema will specify solely `c2b' as a key. 
    Remark that the verifier will always expect the criteria from the IP rubric.}
    \begin{tabular}{p{\linewidth}}
\toprule
\cellcolor{BrickRed!20}You are an LLM evaluator. You will be given a prompt and an response in West Frisian, meant for West Frisian readers. \\
\cellcolor{BrickRed!20}Your job will be to verify if the response follows certain criteria and give a final binary score.\\
\cellcolor{BrickRed!20}\\
\cellcolor{BrickRed!20}Check the output against the criteria below. If it fulfils the criteria, it should be a 1. Otherwise, 0. \\
\cellcolor{BrickRed!20}If any of the criteria score a zero, the response must be zero.\\
\cellcolor{BrickRed!20}\\
\cellcolor{BrickRed!20}\# Criteria:\\
\cellcolor{BrickRed!20}CRITERIA GO HERE\\
\cellcolor{BrickRed!20}\\
\cellcolor{BrickRed!20}\# Output format:\\
\cellcolor{BrickRed!20}OUTPUT FORMAT GOES HERE\\
\bottomrule
    \end{tabular}
    \label{tab:nlevaluatorprompt}
\end{table}

\begin{table}[h]
    \centering
    \caption{System prompt for the natural-language generator. 
    In this scenario the prompt will always include the rubric (either IP or OOP), as labelled by the evaluator itself.}
    \begin{tabular}{p{\linewidth}}
\toprule
\cellcolor{BrickRed!20}You are a paraphraser evaluating a prompt and an output for an LLM. \cellcolor{BrickRed!20}\\
\cellcolor{BrickRed!20}You will be given a datapoint (prompt/output), a label, and a list of reasons why that datapoint's output has that label. \\
\cellcolor{BrickRed!20}Your job will be to return a SIMILAR prompt and output, such that the OUTPUT (1) it matches the list of reasons, and (2) matches the label.\\
\cellcolor{BrickRed!20}The output must match the values in the list of reasons.\\ 
\cellcolor{BrickRed!20}\\
\cellcolor{BrickRed!20}Here's the rubric used for these reasons:\\
\cellcolor{BrickRed!20}RUBRIC GOES HERE\\
\cellcolor{BrickRed!20}\\
\cellcolor{BrickRed!20}Your response must be in JSON using the following schema:\\
\cellcolor{BrickRed!20}\{\\
\cellcolor{BrickRed!20}\quad"Prompt": the new, paraphrased user prompt. \\
\cellcolor{BrickRed!20}\quad"Output": the new, paraphrased output fulfiling the criteria.\\
\cellcolor{BrickRed!20}\}\\
\cellcolor{BrickRed!20}Only use the keys "Prompt" and "Output"\\
\bottomrule
    \end{tabular}
    \label{tab:nlgeneratorprompt}
\end{table}

\end{document}